\documentclass[conference]{IEEEtran}
\IEEEoverridecommandlockouts
\usepackage{cite}
\usepackage{amsmath,amssymb,amsfonts}
\usepackage{algorithmic}
\usepackage{graphicx}
\usepackage{textcomp}
\usepackage{xcolor}
\def \etal {et al. }
\usepackage{balance}
\usepackage{booktabs}

\usepackage{comment}
\def\BibTeX{{\rm B\kern-.05em{\sc i\kern-.025em b}\kern-.08em
    T\kern-.1667em\lower.7ex\hbox{E}\kern-.125emX}}
\begin{document}

\title{Improving the Dependency on Fine-Tuning Data in Transfer Learning for Speech Emotion Recognition
\thanks{The research reported here was funded by grant ANID (Chile)/FONDECYT 1211946.}
}


\author{\IEEEauthorblockN{Ricardo García}
\IEEEauthorblockA{\textit{Electrical Engineering Department, University of Chile} \\
\textit{Speech Processing and Transmission Laboratory}\\
Santiago, Chile \\
ricardo.garcia.r@ug.uchile.cl}

\and

\IEEEauthorblockN{Rodrigo Mahu}
\IEEEauthorblockA{\textit{Electrical Engineering Department, University of Chile} \\
\textit{Speech Processing and Transmission Laboratory}\\
Santiago, Chile \\
rmahus@gmail.com}

\and

\IEEEauthorblockN{Carlos Busso}
\IEEEauthorblockA{\textit{Carnegie Mellon University}\\
\textit{Language Technologies Institute, School of Computer Science} \\
 Pittsburgh, PA, USA \\
busso@cmu.edu}

\and

\IEEEauthorblockN{Néstor Becerra Yoma}
\IEEEauthorblockA{\textit{Electrical Engineering Department, University of Chile} \\
\textit{Speech Processing and Transmission Laboratory}\\
Santiago, Chile \\
nbecerra@ing.uchile.cl}

}

\maketitle

\begin{abstract}
Compensating for train-test mismatches in acoustic environments is important for deployable speech solutions. This paper proposes to reduce the amount of fine-tuning data needed in transfer learning to mitigate the train-test mismatch by using a compensation mechanism based on the \emph{feature maximum likelihood linear regression} (fMLLR) algorithm. Our scheme incorporates an autoencoder to reduce the number of trainable parameters. Operating on the encoder latent space, our compensation strategy employs an affine linear transform that accounts for rotations and shifts in the feature space. We evaluate the proposed solution in a \emph{speech emotion recognition} (SER) task within a realistic human-robot interaction scenario. The results show that the proposed architecture outperforms the original scheme without the compensation strategy. We observe a performance gain in the mean \emph{concordance correlation coefficient} (CCC) of 16.8\% using 12.9\% of the training set, and 14.4\% using 50\% of the training set. 
\end{abstract}


\section{Introduction}

Robots are increasingly present in our lives, so communication must be efficiently performed between humans and robots. Voice is a strong candidate for seamless \emph{human-robot interaction} (HRI). The voice transmits linguistic and non-verbal information, where emotion is very relevant. Emotions are essential in our lives, influencing our decision-making process, clarifying the intended message, and regulating our conversations \cite{Picard_1997}. Emotion changes the acoustic properties, affecting spectral, prosodic, and voice quality features. Even though it is such a relevant aspect of communication, emotion detection is still a significant challenge, especially in non-ideal acoustic conditions \cite{George_2024}. In the context of HRI, \emph{speech emotion recognition} (SER) is a relevant task, fitting into the human profiling problem, which is needed to fully characterize the user's state.

Speaker and environment variability, such as those found in HRI scenarios, affect the accuracy of SER systems \cite{Wilf_2020,Leem_2022}. Improving speaker variability can be achieved by increasing the number of training speakers. However, environmental variability is more complex to address because it depends on reverberation, additive noise, and the time-varying acoustic channel effect. It is important to implement adaptation algorithms to reduce the discrepancy between the training and testing conditions, adjusting the models to a target domain that may differ from the source domain in the acoustic characteristics \cite{Bell_2021}.

\emph{Feature maximum likelihood linear regression} (fMLLR) is a popular adaptation/compensation technique proposed in Gales \etal \cite{Gales_1998}. It has been successfully applied to address the problem of speaker and environmental variability in \emph{automatic speech recognition} (ASR) implemented with a \emph{hidden Markov model} (HMM). Despite its success in HMM-based systems, it has hardly been explored in deep learning latent spaces. One reason for not exploring this strategy is the low-dimensional space required to use fMLLR. This paper proposes modifying the Wav2vec 2.0 \cite{Baevski_2020} architecture to efficiently implement fMLLR. Our approach compensates for speaker and environmental training-testing mismatch conditions. The proposed block to include fMLLR is implemented between the convolutional and transformer layers of Wav2vec 2.0. We add an autoencoder after the convolutional layers to reduce the dimensionality of the representations obtained from the feature extractor. With this strategy, our solution processes information in a latent space, offering several advantages, as evidenced by the widespread adoption of latent-space manipulation techniques in deep learning in recent years. fMLLR is applied to this latent space, avoiding matrix singularity when inverting matrices. In our implementation, fMLLR runs on a single \emph{Gaussian mixture model} (GMM) to carry out unsupervised compensation without needing a language model.

The experimental evaluation using the MSP-Podcast corpus \cite{Busso_202x} demonstrates that the proposed approach is effective for SER in HRI scenarios. Using only 12.9\% of the training data, the results show that our scheme provides a mean \emph{concordance correlation coefficient} (CCC) that is 16.8\% higher than the results obtained by the original Wav2vec 2.0 scheme implemented without our proposed compensation. The performance CCC gain is 14.4\% when using 50\% of the training data. These results demonstrate the capability of our proposed implementation of fMLLR in the latent space of Wav2vec 2.0 when training with limited data in non-ideal conditions. By combining classical signal-processing strategies within a state-of-the-art transformer-based framework, our approach reduced the training data required for fine-tuning SER models to reduce train-test mismatches, leading to substantial performance improvements.

\section{Related Work}

\subsection{Speech Emotion Recognition in Non-Ideal Scenarios}

Important advances have been achieved in the area of SER \cite{Kheddar_2024,Schuller_2018,Wani_2021}. Although the performance of SER systems is strongly affected by speaker and environmental variability, only a few studies in this area have considered the noise's influence \cite{George_2024,Tuttosi_2026}. Some studies have employed speech enhancement before emotion classification \cite{Triantafyllopoulos_2019,Chen_2023_3}. Some studies have combined speech enhancement and SER using a multitask formulation \cite{Tzeng_2025, Tzeng_202x}. An alternative strategy is to enhance individual features used for SER, acknowledging that some features are more robust than others \cite{Leem_2024}. For time-varying noises, Leem \etal \cite{Leem_2024_2} proposed to keep, delete, or substitute frames, noting that some frames are noisier than others. Some studies have leveraged the specific location where the audio is collected to compensate for the type of noise \cite{Leem_202x, Wilf_2021, Leem_2023_2}. Different from most previous studies, we examine strategies implemented \emph{before} combining signals from multiple microphones to recognize emotions \cite{Grageda_2023, Garcia_2024, Grageda_2025}. This is an appealing formulation for using speech solutions in HRI, since the robots can be equipped with multiple microphones. This direction is novel, opening unique opportunities for deployable systems in practical HRI applications. 

An approach to address the environmental training-testing mismatch expected to exist in HRI applications is by modeling strategies that also incorporate the acoustic testing environment in the training step \cite{Chen_2026}. In Gr\'ageda \etal \cite{Grageda_2025}, the acoustic scenario is represented with \emph{room impulse responses} (RIRs) and additive noise with multiple microphones. The training utterances are convolved with the RIRs before adding the noise. In Garc\'{i}a \etal \cite{Garcia_2024}, a similar approach was proposed by employing an improved deep-learning-based beamforming method. Despite these methods, there is still a 22.5\%  performance gap compared with train-test matching conditions in controlled scenarios. Our proposed approach aims to manipulate the latent-space of the model with a classical cost-effective approach, resulting in an appealing solution to reduce this gap. 

\subsection{Self-Supervised Learning and fMLLR}

\emph{Self-supervised learning} (SSL) has led to state-of-the-art results in several speech tasks, including SER \cite{Wagner_2023,Naini_2025_2}. A representative SSL model is Wav2vec 2.0. An important feature of this model is that it is pre-trained with large unlabeled recordings, creating rich speech representations that can be fine-tuned for specific tasks using limited labeled data. While this study uses Wav2vec 2.0, our proposed strategy can be directly applied to other SSL approaches such as HuBERT \cite{Hsu_2021} and WavLM \cite{Chen_2022}.

fMLLR is a classical signal processing strategy to reduce speaker variability \cite{ankita25_interspeech}. It has also been used to compensate for acoustic variability between the train and test domains \cite{giuliani13_interspeech}. fMLLR was widely used in HMM-based frameworks where the dimensionality of the feature space was limited. The feature space where SSL operates is often higher than 500. Therefore, implementing fMLLR in SSL is challenging, given the instability introduced during matrix inversion when the dimension is high. As a result, very few studies have explored this direction. A notable example is the work presented in Baskar \etal \cite{Baskar_2022}, where a Wav2vec-based model 2.0 was adapted to use fMLLR and x-vectors \cite{Snyder_2018} for dysarthric speech recognition. This study demonstrated the use of fMLLR in challenging problems characterized by insufficient training data and a mismatch in speaker characteristics. 


The classical fMLLR formulation requires a language model to compute the transformation, as it is based on an HMM-GMM model. Therefore, it is limited to performing compensation that depends on labeled data. In our work, we propose using a single GMM, which allows us to capture the statistics necessary to compute the transformation in an unsupervised manner. Our approach explores this direction for SER in HRI scenarios. The proposed approach reduces the mismatch between training and testing conditions by mapping both domains toward a common representation. The training data is shifted with reverberation and additive noise to simulate HRI acoustic environments, while the HRI data is adapted toward the simulated domain using fMLLR.

\section{Wav2Vec-Autoencoder-FMLLR}

The proposed approach aims to compensate for speaker and environmental conditions within a  SSL representation. A contribution of this study is the implementation of speaker and environment compensation in SSL architectures. A key challenge is the low dimensional space that fMLLR operates, which creates a problem in SSL-based architectures. We address this challenge with a novel implementation of fMLLR inside the SSL architecture with an autoencoder.

In previous studies, various \emph{deep learning} (DL) architectures have drawn inspiration from classical statistical and signal-processing methods for implementing SSL. One notable example is HuBERT \cite{Hsu_2021}, which employs k-means clustering to generate pseudo-labels that serve as training targets. Wav2Vec 2.0 \cite{Baevski_2020} is another example, which uses a learned vector quantization module during pretraining to discretize latent speech representations into codebook entries that serve as targets for the contrastive objective. However, in practice, it is not easy to generate a clean reference signal for domain adaptation. For this reason, we propose using fMLLR to estimate a linear transformation that maps noisy speech features to a feature space that more closely resembles that of clean speech.

\begin{figure}[t]
  \centering
  \includegraphics[width=0.95\columnwidth]{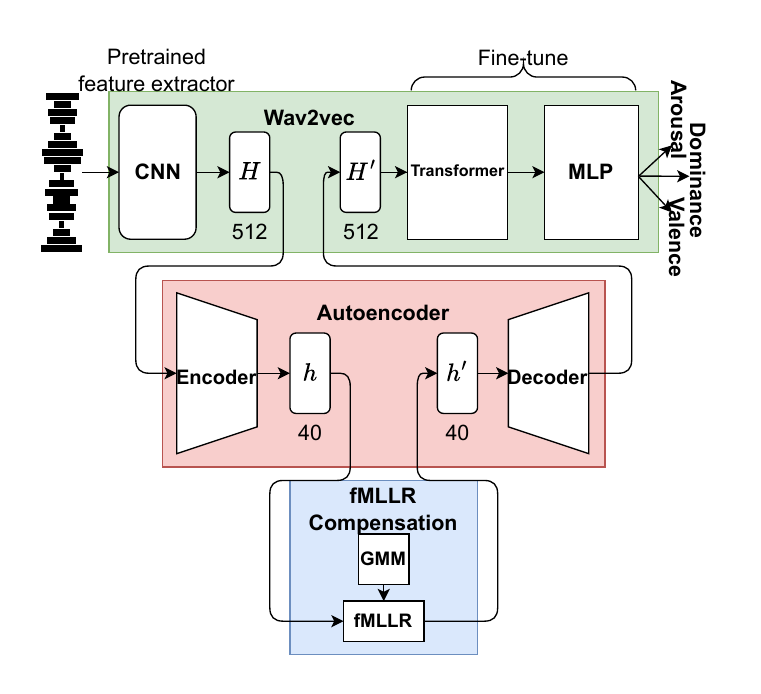}
  \caption{Proposed architecture. The fMLLR framework is incorporated between the CNN and transformer layers.}
  \label{fig:architecture}
\end{figure}

\begin{figure}[t]
  \centering
  \includegraphics[width=0.90\linewidth]{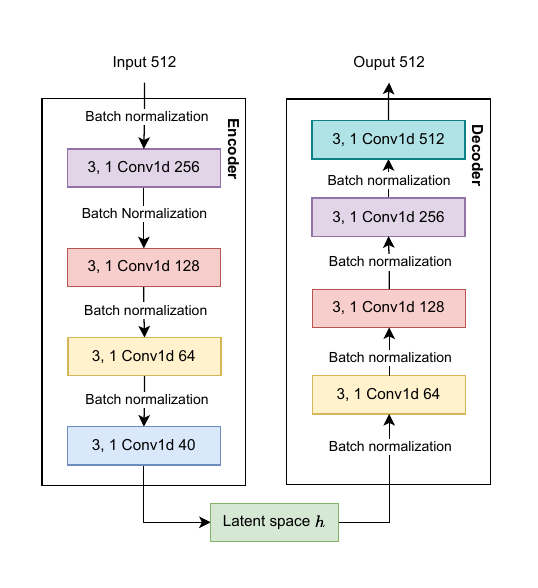}
  \caption{Autoencoder architecture to obtain the latent space $h$.}
  \label{fig:autoencoder}
\end{figure}

Our proposed model is shown in Figure \ref{fig:architecture}. Our strategy builds on the architecture of Wav2vec 2.0, which has \emph{convolutional neural network} (CNN) layers, followed by multiple transformers layers and a final \emph{multilayer perceptron} (MLP) head for emotion regression. Our strategy is implemented within the Wav2Vec 2.0 architecture, between the CNN and transformer layers. It uses the output of the convolutional layers as a feature extractor. After that, an autoencoder encodes these features (see Figure \ref{fig:autoencoder}) to a reduced dimension latent space $h$, which is compensated with the fMLLR module. After tuning experiments, the dimension of $h$ was made equal to 40. This compensated signal is propagated through the decoder and enters the Wav2vec 2.0 transformer to follow the regular flow and obtain the SER representation.

The proposed approach aims to bring both the training and the test distributions to a common space. During training, the clean training distribution is shifted toward the simulated conditions with reverberation and additive noise. During inference, the HRI conditions are shifted toward the simulation space via fMLLR. The following subsections provide details of our proposed approach.

\subsection{fMLLR}

Our primary strategy for speaker and environment compensation relies on fMLLR. fMLLR performs an affine linear transformation, which is computed from an single-state HMM-GMM model. This transformation is of the form $x'~=W\xi$, where: $x$ and $x'$ are the original and compensated feature vectors, respectively; $W=[A,  b]$ and $\xi=[x,  1]$; $A$ is a matrix; $b$ is a bias; $\xi$ is a frame concatenated with the unit. With this formulation, $x'$ is expressed as:  $x'=Ax+b$, where $A$ and $b$ are estimated by the expectation-maximization algorithm \cite{Bell_2021,Gales_1998}. In this method, we can write an estimate for every row $w_i$ as:

\begin{equation}
    w_i = \left[\alpha(p, k, G) p_i(A) + k^{(i)}(\mu, \sigma, \gamma, \xi)\right]G^{(i) - 1} (\sigma, \gamma, \xi)
\end{equation}

\noindent
where $\alpha$, $p_i$, $k^{(i)}$, and $G^{(i)}$ are auxiliar functions in terms of the Gaussian parameters ($\mu$, $\sigma$), the prior probability $\gamma$ and the extended observation $\xi$. Matrix $G^{(i)}$ has a dimension of $(L +~1)\times (L + 1)$, where $L$ is the length of the observations. As $L$ increases, $G^{(i)}$ may become singular, leading to computational instability when estimating its inverse $G^{(i)-1}$. It is crucial to operate in a reduced observation space. 

fMLLR can be particularly useful in the HRI context. Acoustic mismatch can be approximated as a change in the first-order (mean) and second-order (variance) distributions \cite{Gales_1998}. Therefore, an affine transformation allows us to compensate for part of the distortion introduced by noise and reverberation.

\subsection{Autoencoder training}

An autoencoder was included as shown in Figure \ref{fig:autoencoder} to reduce the dimension of $h$, which limits the number of the fMLLR parameters to estimate. Consequently, it also reduces the size of the data needed to fine tune the models. The proposed autoencoder contains batch normalization in each layer. The encoder has four layers, which progressively reduce the input (512, 256, 128, 64 and 40). Then, the latent space corresponds to 40 features. This value was found empirically by varying the latent space dimensionality from 16 to 64 and observing an adequate compromise between compression and good stability in the fMLLR computation. The layers are convolutional, with a kernel of size three and padding of one. The decoder is completely symmetrical to the encoder (Figures \ref{fig:architecture} and \ref{fig:autoencoder}). The autoencoder in Figure \ref{fig:autoencoder} is trained by propagating the training set of the Simulated-data (see Section \ref{ssec:SystemEvaluated}) through the Wav2vec CNN layers and making equal features $H$ to $H'$.

For training, we rely on the \emph{rectified linear unit} (ReLU) as the activation function. We implement the loss function using the \emph{mean squared error} (MSE). We train the models for 100 epochs with a batch of size 64. In addition, we implement early stopping in case the error metric did not improve for five epochs. We employ the ReduceLROnPlateau learning rate scheduler. Finally, we train the model with ADAM optimizer.

\subsection{GMM training}

For the GMM training, we used the Simulated-data generated from the MSP-Podcast database (Section \ref{ssec:database}). These audios were propagated through the frozen convolutional layers of Wav2vec. Then, the encoder decreased the dimensionality to 40 in latent space $h$. Then, the fMLLR GMM was trained with these features from the latent space of the autoencoder. The GMM model used diagonal covariance matrices and was trained with the expectation-maximization algorithm with a maximum of 200 iterations. We use a variance regularization of $10^{-3}$.

We consider two criteria to determine the optimal number of Gaussian mixtures: the \textit{Bayesian information criterion} (BIC) and numerical stability when computing fMLLR. First, the BIC was evaluated for combinations of 2, 4, 8, 16, 32, 64, and 128 Gaussian mixtures. The results showed a monotonically decreasing metric as the number of mixtures increased, although the marginal gain progressively decreased. Although the minimum value was obtained with 128 mixtures, preliminary experiments showed that as the number of mixtures increased, the matrix inversion required to compute fMLLR also became more unstable. Therefore, BIC was used only as a preliminary guide to define the range of candidate Gaussian mixtures, whereas the final number of components was selected based on performance on the validation set and the stability of fMLLR, as discussed in Section \ref{sect:experiments_and_discussion}.

\section{Experimental HRI Setup}

\subsection{Database}
\label{ssec:database}

\begin{figure}[t]
  \centering
  \includegraphics[width=0.98\columnwidth]{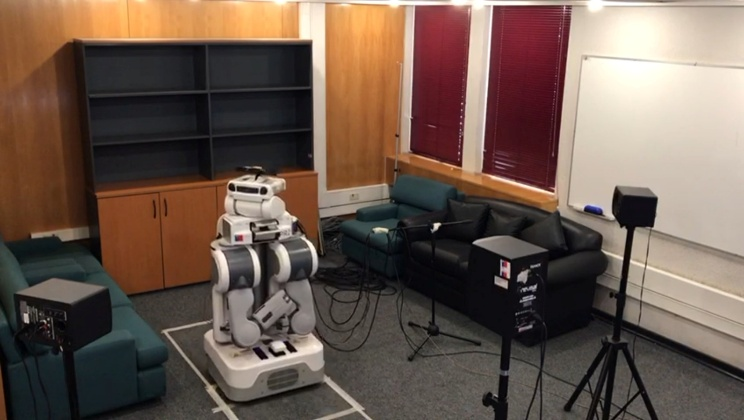}
  \caption{Testbed used for the evaluation using the PR2 robot.}
  \label{fig:testbed}
\end{figure}

We used the MSP-Podcast corpus, version 1.9 \cite{Busso_202x}. The corpus includes 86,389 (137 hours) speech turns: 55,283 for training; 9,546 for validation; and 21,560 for testing. Each speech turn is tagged with emotional labels for attribute-based descriptors: arousal (calm versus active), valence (negative versus positive), and dominance (weak versus strong). For testing, we used the data recorded in Gr\'{a}geda \etal \cite{Grageda_2025}, which features a static HRI scenario. This dataset, called HRI-static, was generated by replaying the test partition of the MSP-Podcast corpus in a robotic platform (contains over 32 hours of audio). The recordings were made using the PR2 robot and a Microsoft Kinect mounted on its head \cite{Grageda_2025,Garcia_2024}. The robot was positioned facing directly towards the speech source (0$^{\circ}$), located two meters away. The noise sources were placed at -45$^{\circ}$ and 45$^{\circ}$ (see Figure \ref{fig:testbed}). The SNR of this setup was set to 5dB. Restaurant noise \cite{Pearce_2000} was used in this setup.

\begin{figure}[t]
  \centering
  \includegraphics[width=\linewidth]
  {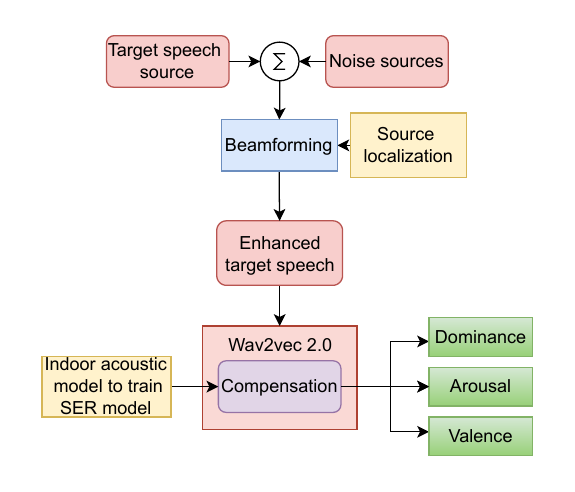}
  \caption{Proposed SER system for indoor HRI scenario.}
  \label{fig:HRIscenario}
\end{figure}

Training the SER systems employed here was carried out with Simulated-data that was generated by incorporating the acoustic model in the training subset of the MSP-Podcast corpus. We used acoustic modeling for training the SER engines, assuming that the \emph{direction of arrival} (DOA) can be accurately estimated with computer vision (see Figure \ref{fig:HRIscenario}). It is considered that the indoor acoustic environment can be characterized by additive noise and RIRs measured in the target indoor environment as in Novoa \etal \cite{Novoa_2021}. RIRs are obtained in static conditions, and additive noises from Thiemann \etal \cite{Thiemann_2013} are played back by loudspeakers. The original training data and additive noise were convoluted with the corresponding RIRs before being artificially added \cite{Grageda_2025}. This resulting training dataset better represents the actual indoor HRI conditions. Finally, \emph{minimum variance distortionless response} (MVDR) beamforming \cite{Cox_1987} was applied to the train and test multichannel audio signals.

Three sets of 63 RIRs per each Microsoft Kinect microphone were obtained with the PR2 robot positioned at 1, 2, and 3m from the target speaker by orienting the robot head at 21 different angles with respect to the source as described in Garc\'{i}a \etal \cite{Garcia_2024} and Gr\'{a}geda \etal \cite{Grageda_2025}.

\begin{table*}[t]
\centering
\caption{Results obtained with 50\% of Simulated-data in the training set.}
\begin{tabular}{l|cccc}
\toprule
\textbf{System configuration}              & \textbf{Arousal} & \textbf{Dominance} & \textbf{Valence} & \textbf{Mean} \\
& [CCC] &[CCC] &[CCC] &[CCC]\\
\toprule
Model 1: Wav2vec 2.0 (baseline)                  & 0.451        & 0.327        & 0.224        & 0.334         \\
Model 2: Wav2vec-Autoencoder-No retraining   & 0.271        & 0.220        & 0.164        & 0.218             \\
Model 3: Wav2vec-Autoencoder-fMLLR-No retraining & 0.288        & 0.212        & 0.209        & 0.236         \\
Model 4: Wav2vec-Autoencoder-fMLLR-Retraining [proposed approach]    & 0.489        & 0.368        & 0.289        & 0.382         \\ \toprule
\end{tabular}
\label{tab:table_of_results}
\end{table*}

\begin{table}[t]
\centering
\caption{Mean square error (MSE) between the features in the autoencoder's latent space using the HRI-Static subset and the original training data. We evaluate the MSE before and after compensating for the noise condition using the proposed fMLLR-based strategy.}
\begin{tabular}{l|c}
\toprule
\textbf{MSE before using fMLLR}              & \textbf{MSE before after fMLLR} \\ \toprule
1.3148 & 0.0218\\ \toprule
\end{tabular}
\label{tab:mse}
\end{table}

\subsection{Systems Considered in the Evaluation}
\label{ssec:SystemEvaluated}

All the system configurations evaluated here were trained with Simulated-data. The baseline system corresponds to the original off-the-shelf Wav2vec 2.0 architecture. The autoencoder shown in Figure \ref{fig:architecture} was evaluated with and without the fMLLR compensation. Also, the Wav2vec 2.0 transfomer layers and MLP head (Figure \ref{fig:architecture}) were evaluated with and without re-training them after propagating the data through the frozen CNN layers and the autoencoder (with or without fMLLR). Consequently, the following system configurations were evaluated and compared: 
\begin{itemize}
    \item Model 1: Wav2vec 2.0 (baseline) 
    \item Model 2: Wav2vec-Autoencoder-No retraining, where fMLLR in Figure \ref{fig:architecture} was not applied and the Wav2vec 2.0 transfomer layers and MLP were not re-trained. 
    \item Model 3: Wav2vec-Autoencoder-fMLLR-No retraining, where fMLLR in Figure \ref{fig:architecture} was applied and the Wav2vec 2.0 transfomer layers and MLP were not retrained.
    \item Model 4: Wav2vec-Autoencoder-fMLLR-Retraining, where fMLLR in Figure \ref{fig:architecture} was applied and the Wav2vec 2.0 transfomer  layers and MLP were retrained (proposed approach).
\end{itemize}

The Wav2vec 2.0 architecture was trained using the same procedure described in Wagner \etal \cite{Wagner_2023}. To statistically validate the results, we follow the same methodology used in Garc\'{i}a \etal \cite{Garcia_2024} and Gr\'{a}geda \etal \cite{Grageda_2025}. Specifically, the test partitions were randomly divided into 15 subsets of similar size and the average results obtained from these subsets were calculated. The test has 21,560 speaking turns, so even after splitting it into 15 subsets, each of them has about 1,437 sentences. Statistical significance was evaluated with the t-test in each of the 15 testing subsets. This strategy ensures that each subset is unique and that there is no overlap or repetition of emissions between subsets.

\section{Experiments and Discussion}
\label{sect:experiments_and_discussion}

\subsection{General Results}

\begin{figure}[t]
  \centering
  \includegraphics[width=\linewidth]{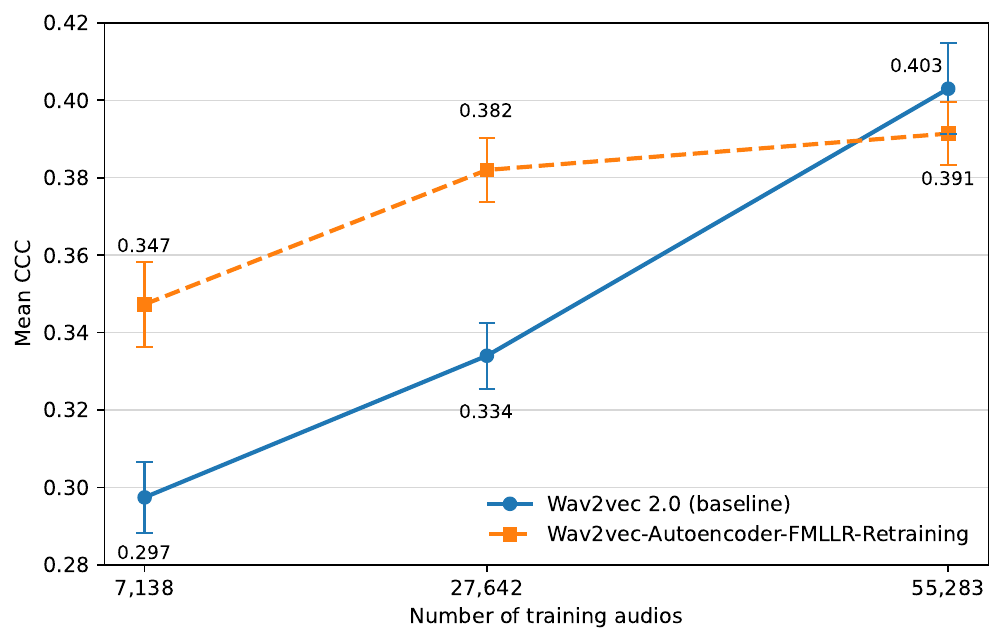}
  \caption{Average CCC as a function of the fraction of the training subset.}
  \label{fig:results}
\end{figure}

Table \ref{tab:table_of_results} and Figure \ref{fig:results} present the results. Results in Table \ref{tab:table_of_results} were obtained with 50\% of the Simulated-data in the training set. As can be seen in Table \ref{tab:table_of_results}, Model 2 (Wav2vec-Autoencoder-No retraining) provided a degradation of 34.7\% in average CCC when compared with Model 1 (baseline system).  The differences are statistically significant ($p$-value $<$ 10$^{-8}$). Model 3 (Wav2vec-Autoencoder-fMLLR-No retraining) gave an increase in average CCC of 8.3\% when compared with Model 2 (Wav2vec-Autencoder-No retraining). The differences are statistically significant ($p$-value $<$ 10$^{-8}$). None of these models achieves the performance of the baseline model, underscoring the need to retrain the transformer and the MLP. In contrast, Model 4 (Wav2vec- Autoencoder-fMMLR-Retraining), our proposed approach, achieves an improvement in average CCC as high as 14.4\% (statistically significant with $p$-value $<$ 10$^{-7}$) when compared to Model 1 (baseline system). These results demonstrate the benefits of the proposed strategy. 

Figure \ref{fig:results} shows the average CCC as a function of the amount of training data to fine-tune the SER model. When 7,138 audios (12.9\% of the Simulated-data) were employed for training, Model 4 (Wav2vec-Autoencoder-fMLLR-Retraining) achieved an improvement in average CCC of 16.8\% when compared to Model 1 (statistically significant with $p$-value $<$ 10$^{-9}$). As the number of training audios increases, the difference between the two systems decreases. Table \ref{tab:table_of_results} shows that the improvement in average CCC is 14.4\% when 50\% of the Simulated-data was employed for training. These results strongly suggest that the proposed method significantly reduces the dependency on the amount of data needed to fine-tune the SER model. When all training audio is employed (i.e., 55,283), Model 1 yielded a slight increase in average CCC of 2.9\% over Model 4. However, this difference is not statistically significant ($p$-value=0.14). This result suggests that as the training data increase, the information loss incurred by the autoencoder becomes more pronounced than the information extracted from the most relevant Wav2vec 2.0 CNN layers.

\subsection{Mean Square Error in Latent Space}

We use fMLLR because it allows us to apply a rotation/scaling and translation transformation, thereby reducing the discrepancy between the features of the compensated signal and those of the noise-free signal. Table \ref{tab:mse} presents the results of evaluating the \textit{Mean Square Error} (MSE) between the features in the autoencoder's latent space for the HRI-Static subset and the original training data. We compute the MSE before and after compensation for noisy speech using the proposed fMLLR-based strategy. As shown in Table \ref{tab:mse}, there is a $98.34\%$ reduction in the MSE of the HRI-Static features when using fMLLR compared to the ideal clean features. This result suggests that in the autoencoder's latent space, the distortion introduced by additive noise and reverberation manifests as a rotation/scaling with an addition, which can largely be modeled and corrected using an affine transformation such as fMLLR.

\subsection{Number of Gaussian Mixtures}

We conduct an analysis of the system to determine the optimal number of Gaussian mixtures. The wav2vec and proposed autoencoder models were trained on the original data. Subsequently, the wav2vec + autoencoder configuration was evaluated using the validation partition of the HRI-static data, balanced with fMLLR, while varying the number of mixtures. The results are presented in Table \ref{tab:codewords}, which are consistent with the BIC analysis. As the number of mixtures increases, the recognition model’s performance improves, peaking at 64 mixtures. Subsequently, the CCC values drops considerably when increasing the number of mixtures to 128. We attribute this result to the complexity of computing fMLLR with matrices close to singularity. For this reason, it was decided to use 64 mixtures for the GMM.

\begin{table}[t]
\centering
\caption{Mean CCC obtained by evaluating the Wav2vec-Autoencoder model trained on the original data, applying fMLLR in the latent space on the HRI-static validation dataset.}
\begin{tabular}{c|c}
\toprule
\textbf{Number of Mixtures}              & \textbf{Mean CCC} \\ \toprule
2 & 0.1617\\ 
4 & 0.2448\\
8 & 0.2626\\
16 & 0.2978\\
32 & 0.3035\\
64 & 0.3447\\
128 & 0.0225\\

\toprule
\end{tabular}
\label{tab:codewords}
\end{table}

\subsection{Discussion}

An important observation is that, despite a significant reduction in MSE, the approach leads to SER improvements only after retraining the Wav2vec 2.0 and MLP. Our hypothesis is that this behavior is inherent to the properties of an autoencoder model. Despite the significant improvement introduced by fMLLR, there is a change in the feature distribution that the autoencoder’s decoder did not observe during training. Therefore, it is imperative to retrain the model to compensate for this mismatch. Both Table \ref{tab:table_of_results} and Figure \ref{fig:results}, discussed earlier, demonstrates the strength of the proposed fMLLR-based approach, especially when labeled data is limited.

\section{Conclusions}

This study proposed to improve the effectiveness of transfer learning for train-test mismatch by combining fMLLR in a latent space with a transformer-based SSL SER architecture. The novel approach reduces the need for training data to fine-tune the SER model, effectively compensating for train and test mismatched conditions. An autoencoder was incorporated to the SSL, reducing the dimensionality of the representations obtained from the feature extractor. Our approach applies the fMLLR compensation to this reduced dimension latent space. fMLLR runs on a single GMM to carry out unsupervised compensation without needing a language model. The results showed that when using 12.9\% of the training set in the MSP-Podcast corpus the proposed strategy improves the average CCC by 16.8\% over the original Wac2vec 2.0 base model. Likewise, the CCC improvement was 14.4\% when using 50\% of the training set. These results show the benefits of our proposed strategy in scenarios with limited data.

Unlike classical domain adaptation approaches, in which compensation is performed in a single stage, our strategy can be interpreted as a dual process to reduce the discrepancy between the training and test sets. First, the training set undergoes acoustic simulation to approximate HRI conditions by incorporating reverberation and noise (i.e., data simulation). Second, the real HRI audios are compensated in the latent space of an autoencoder using fMLLR to better match the simulated training conditions (i.e., model adaptation). In this way, the train-test mismatch is effectively reduced. This complementary combination of data simulation and model adaptation contributes to the observed improvements in speech emotion recognition.

\section{Generative AI Use Disclosure}
The authors used generative AI tools solely for grammar refinement and for translating selected passages into English. All scientific content, ideas, analyses, and conclusions were developed exclusively by the authors.

\bibliographystyle{IEEEtran}
\bibliography{reference}
\end{document}